\documentclass[onecolumn,pra,nofootinbib]{revtex4-2}

\usepackage[dvips]{graphicx} 
\usepackage{amsfonts}
\usepackage{amssymb}
\usepackage{amscd}
\usepackage{amsmath}
\usepackage{enumerate}
\usepackage{epsfig}
\usepackage{subfigure}
\usepackage{subfloat}
\usepackage{xcolor}
\usepackage{amsthm}
\usepackage{physics}
\usepackage[most]{tcolorbox}
\usepackage{ulem}
\usepackage{tikz}
\usetikzlibrary{quantikz}
\usetikzlibrary{backgrounds}
\usepackage{tabularx}
\usepackage{float}
\usepackage{appendix}
\usepackage{makecell}
\usepackage[ruled,vlined]{algorithm2e}
\usepackage{algpseudocode}
\usepackage{dsfont}
\usepackage[colorlinks, linkcolor=red, anchorcolor=blue, citecolor=green]{hyperref}
\usepackage{MnSymbol}
\newtcolorbox[auto counter]{mybox}[2]{
	enhanced,
	breakable,
	label=#1,
	colback=blue!5!white,
	colframe=blue!75!black,
	fonttitle=\bfseries,
	title=Box \thetcbcounter: #2
}

\newcommand{\lket}[1]{\vert #1 \rangle\!\rangle}
\newcommand{\lbra}[1]{\langle\!\langle #1 \vert}
\newcommand{\lbraket}[2]{\langle\!\langle #1 \vert #2 \rangle\!\rangle}

\newcommand{\tra}[1]{\mathrm{tr}(#1)}
\newcommand{\Pauli}{\mathbb{P}}
\newcommand{\NU}{\tilde{\mathcal{U}}}

\newcommand{\comments}[1]{}

\begin{document}

\title{Benchmarking non-Clifford gates using only Pauli twirling group}
\author{Han Ye, Guoding Liu, and Xiongfeng Ma}
\email{xma@tsinghua.edu.cn}
\affiliation{Center for Quantum Information, Institute for Interdisciplinary Information Sciences, Tsinghua University, Beijing, 100084 China}

\begin{abstract}
Quantum gate benchmarking is unavoidably influenced by state preparation and measurement errors. Randomized benchmarking addresses this challenge by employing group twirling to regularize the noise channel, then provides a characterization of quantum channels that is robust to these errors through exponential fittings. In practice, local twirling gates are preferred due to their high fidelity and experimental feasibility. However, while existing RB methods leveraging local twirling gates are effective for benchmarking Clifford gates, they face fundamental challenges in benchmarking non-Clifford gates. In this work, we solve this problem by introducing the Pauli Transfer Character Benchmarking. This protocol estimates the Pauli transfer matrix elements for a quantum channel using only local Pauli operations. Building on this protocol, we develop a fidelity benchmarking method for non-Clifford gates $U$ satisfying $U^2=I$. We validate the feasibility of our protocol through numerical simulations applied to Toffoli gates as a concrete example.
\end{abstract}

\maketitle

\section{Introduction}\label{sec:intro}
The accurate benchmarking of quantum processes~\cite{Eisert2020benchmarking} is a critical technological prerequisite for realizing high-precision quantum gates~\cite{Monz2009Toffoli,Kim2022iToffoli,Evered2023CCZ,Nguyen2024CCZ,Wang2024Toffoli,Fan2025paraCZ} and, consequently, for the development of quantum error correction and fault-tolerant quantum computation~\cite{Egan2021corrected,Gong2021correcting,Postler2022tolerant,Zhao2022Correcting,Acharya2023Suppressing,Acharya2025threshold}. A variety of protocols have been developed to characterize quantum operations, ranging from quantum process tomography~\cite{Chuang01111997}, to direct fidelity estimation~\cite{Flammia2011DFE} and randomized benchmarking (RB)~\cite{Knill2008RB, Emerson2011prlRB, Emerson2012praRB,Helsen2022RBFramework}. A pivotal practical consideration in applying these methods is their susceptibility to state preparation and measurement (SPAM) errors~\cite{Nielsen2021gatesettomography}, which can significantly distort the inferred characteristics of the target quantum process. Consequently, the development of benchmarking techniques that are robust to SPAM errors has emerged as a critical research direction~\cite{Helsen2022RBFramework}.

Among these, RB and its variants~\cite{magesan2012efficient,Erhard2019CB,helsen2019new,Helsen2022RBFramework} provide SPAM error-free methodologies to estimate the fidelity of quantum processes. More recent innovations have further expanded the capabilities of this framework by integrating RB with shadow tomography~\cite{helsen2023estimating,wang2024robust}, enabling the estimation of additional gate properties beyond gate fidelity.

The core techniques of RB include employing group twirling to regularize the noise channel and extract information in a SPAM error-free manner through exponential fittings. The group twirling technique typically requires the application of additional quantum gates, termed twirling gates. Among the various choices of twirling gates, local gates, such as Pauli or single-qubit Clifford gates, are preferred for several reasons. Firstly, the gate-independent noise assumption of the twirling gates provides theoretical tractability and robustness~\cite{Emerson2012praRB}, while local gates are closer to satisfying the gate-independent noise assumption than multi-qubit gates. Secondly, their implementation involves only single-qubit gates, which avoids the overhead for complex multi-qubit gate compilation and improves the experimental feasibility. Finally, given that many quantum hardware platforms natively support high-fidelity local gates~\cite{Smith2025Gates,Acharya2025threshold,Gao2025Zuchongzhi3,Sales2025magic}, the utilization of such high-fidelity twirling operations leads to more accurate noise estimation~\cite{magesan2012efficient}. For these reasons, the use of local twirling gates is highly advantageous.

Hence, a central question arises: what kind of gates can be reliably characterized using only local twirling gates? Previous research has demonstrated that Pauli twirling gates can effectively benchmark the noise channel of Clifford gates, up to local gauge transformations~\cite{carignan2015characterizing,Erhard2019CB,zhang2022scalable,Chen2023learnability}. However, for generic multi-qubit non-Clifford gates, progress remains limited. Existing approaches for benchmarking non-Clifford gates typically require multi-qubit twirling groups~\cite{cross2016scalable,garion2021experimental,liu2024group}. Moreover, it has been shown that in a specific framework, local twirling gates cannot effectively twirl the noise channel of multi-qubit controlled phase gates~\cite{liu2024group}. These results suggest that designing an RB protocol for multi-qubit non-Clifford gates using only local twirling gates is exceptionally challenging. On the other hand, non-Clifford gates are indispensable for universal quantum computation~\cite{gottesman1998heisenberg,aaronson2004improved}, underscoring the practical need for accurately characterizing their performance.

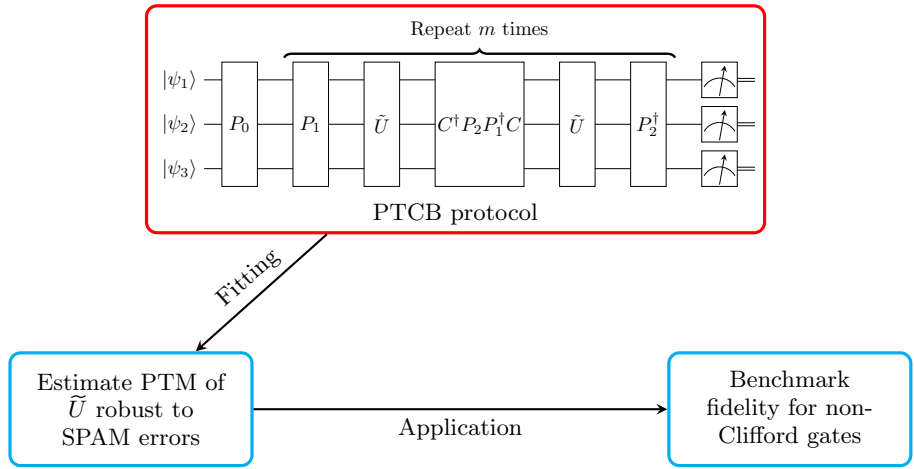
\begin{figure}[htbp!]
    \centering
    \begin{tikzpicture}[
        process/.style={draw=cyan,rectangle,very thick, rounded corners,minimum height=1.5cm,text width=3cm,align=center,text=black},
		startstop/.style={draw=red,rectangle,very thick, rounded corners, minimum width=3cm, minimum height=1cm,align=center},
		arrow/.style={thick,->,>=stealth},
		node distance=2.2cm]
        \node (start) [startstop] {{\includegraphics[width=8cm]{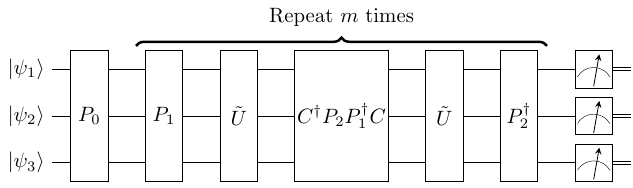}}\\
        PTCB protocol};
        \node (result) [process, below left=of start, xshift=3cm] {Estimate PTM of\\$\widetilde{U}$ robust to SPAM errors};
        \node (application) [process, right= of result, xshift=3.3cm] {Benchmark\\fidelity for non-Clifford gates};
        \draw [arrow] (start) -- node[midway,above,sloped] {Fitting} (result);
        \draw [arrow] (result) -- node[midway,below] {Application} (application);
    \end{tikzpicture}
    \caption{The schematic diagram of the main contributions of this work. The quantum circuit effectively illustrates the core concept of the PTCB protocol. While the introduction of the Clifford gate $C$ enables access to arbitrary PTM elements of $\tilde{U}$, it simultaneously introduces uncontrolled additional noise. To mitigate this challenge, we incorporate the conjugate gate $C^\dagger$ and combine $C$, $C^\dagger$, and $P_2 P_1^\dagger$ into a single Pauli gate, thereby circumventing the physical implementation of Clifford gates altogether.}
    \label{fig:flowchart}
\end{figure}

In this work, we tackle the above problem by proposing the Pauli Transfer Character Benchmarking (PTCB), a protocol designed to estimate the product of any two symmetric elements of the Pauli transfer matrix (PTM) for a quantum channel utilizing only the Pauli twirling group and Pauli-basis SPAM. Building upon the PTCB module, we then propose a benchmarking protocol for non-Clifford gates $U$ satisfying $U^2 = I$ (a constraint that may be further relaxed under appropriate noise assumptions). Whereas conventional group twirling frameworks aim to twirl the noise channel directly, our protocol instead constructs the twirled channel as the entire target gate combined with a Clifford gate. Through the appropriate selection of the Clifford gate, any PTM element can be accessed, especially the off-diagonal ones. A similar approach of using Clifford gates to probe off-diagonal PTM elements appears in~\cite{francca2021efficient}. Our method is distinguished in its use of a virtual conjugated Clifford pair, which is not physically implemented, ensuring that all operations except the target gate itself remain local. We further validate the feasibility of our protocol through numerical simulations applied to Toffoli gates as a concrete example. The main contributions of our work is summarized in Fig.~\ref{fig:flowchart}.

The structure of this article is as follows. In Sec.~\ref{sec:pre}, we review necessary background on group twirling and benchmarking. In Sec.~\ref{sec:ptcb}, we detail the PTCB protocol. In Sec.~\ref{sec:simulation}, we present numerical simulations for Toffoli gates. Finally, we conclude in Sec.~\ref{sec:discussion}.

\section{Preliminaries}\label{sec:pre}
\subsection{Pauli-Liouville representation}
The $n$-qubit Pauli group and its normalized counterpart, modulo global phases in $\{\pm1, \pm i\}$, are defined as:
\begin{equation}
    \begin{gathered}
        \Pauli_n=\{I,X,Y,Z\}^{\otimes n},\\
        \Pauli'_n=\{\sigma_P=\frac{P}{\sqrt{d}}|P \in \Pauli_n\},
    \end{gathered}
\end{equation}
where $d = 2^n$ denotes the Hilbert space dimension. For brevity, we omit the subscript $n$ when unambiguous. The normalized Pauli group satisfies $\tra{\sigma_P^\dag \sigma_Q} = \delta_{PQ}$, thus forming an orthonormal basis for the space of $d$-dimensional operators. Any operator $O$ admits the expansion:
\begin{equation}\label{eq:representation}
    O=\sum_{P\in\Pauli}\tra{\sigma_P^\dag O}\sigma_P.
\end{equation}
In Pauli-Liouville representation, Eq.~\eqref{eq:representation} is equivalent to:
\begin{equation}
    \lket{O}=\sum_{P\in\Pauli}\lbraket{\sigma_P}{O}\lket{\sigma_P},
\end{equation}
where $\lket{O}$ is a $d^2$-dimensional vector, $\lket{\sigma_P}$ constitutes an orthonormal basis, and $\lbraket{\sigma}{O} = \tra{\sigma^\dag O}$ defines the inner product.

A quantum channel $\Lambda$ can be represented by its Pauli transfer matrix (PTM), defined as:
\begin{equation}\label{eq:PTM def}
    \Lambda_{PQ}=\lbraket{\sigma_P}{\Lambda(\sigma_Q)}.
\end{equation}
By linearity, the action of $\Lambda$ corresponds to matrix multiplication:
\begin{equation}
\lket{\Lambda(O)}=\Lambda\lket{O}.
\end{equation}
Here, $\Lambda$ denotes both the quantum process and its PTM.

In this article, we adopt the following notational conventions. Unitary operations are denoted by standard fonts (e.g., $P$ for a Pauli unitary), sets of unitaries by blackboard bold (e.g., $\Pauli$ for the Pauli group), and PTMs by calligraphic fonts (e.g., $\mathcal{P}$ for the PTM of $P$). Noisy channels are indicated by $\tilde{\cdot}$. Specifically, the target gate is denoted by $U$.

\subsection{Quantum channel fidelity}
Various measures quantify the fidelity of a quantum channel $\Lambda$, that is, how close it is to the identity. Benchmarking aims to estimate such a measure experimentally. The average fidelity and process fidelity are defined as:
\begin{equation}\label{eq:fidelity}
    \begin{gathered}
        F_{\textrm{avg}}(\Lambda)=\int \mathrm{d}\psi\ \tra{\ketbra{\psi}\Lambda(\ketbra{\psi})},\\
        F(\Lambda)=\frac{1}{d^2}\tra{\Lambda}=\frac{1}{d^2}\sum_{P\in\Pauli}\Lambda_{PP},
    \end{gathered}
\end{equation}
where $\mathrm{d}\psi$ denotes integration over the Haar measure. The two measures are related by \cite{horodecki1999general}:
\begin{equation}\label{eq:relation}
    F_{\textrm{avg}}=\frac{d\cdot F+1}{d+1}.
\end{equation}
Henceforth, all fidelities refer to process fidelities. For a unitary $V$ actually implemented as $\tilde{\mathcal{V}}=\mathcal{V}\Lambda$ or $\tilde{\mathcal{V}}=\Lambda\mathcal{V}$, where $\mathcal{V}$ is the ideal gate and $\Lambda$ is the noise, the fidelity of $\tilde{\mathcal{V}}$ refers $F(\Lambda)$.

\subsection{Group twirling}
For a quantum channel $\Lambda$ and a unitary group $\mathbb{G}$, the twirled channel is defined as:
\begin{equation}
\Lambda_\mathbb{G}=\mathbb{E}_{G\in\mathbb{G}}\mathcal{G}^\dag\Lambda\mathcal{G}.
\end{equation}
From Eq.~\eqref{eq:fidelity}, it follows that:
\begin{equation}
    \begin{aligned}
        F(\Lambda_\mathbb{G})&=\frac{1}{d^2}\tra{\mathbb{E}_{G\in\mathbb{G}}\mathcal{G}^\dag\Lambda\mathcal{G}}\\
        &=\frac{1}{d^2}\mathbb{E}_{G\in\mathbb{G}}\tra{\mathcal{G}^\dag\Lambda\mathcal{G}}\\
        &=\frac{1}{d^2}\mathbb{E}_{G\in\mathbb{G}}\tra{\Lambda}\\
        &=F(\Lambda).
    \end{aligned}
\end{equation}

Before discussing specific twirling groups, we elaborate on the gate-independent noise assumption and the issue of gate compilation mentioned in Sec.~\ref{sec:intro}, which arise naturally from the operational distinctions between single- and multi-qubit gates. In general, the implementation of multi-qubit gates requires compiling them into sequences of native single- and two-qubit gates. This compilation process introduces two significant challenges. First, the required classical and quantum resources proliferate with the number of qubits. Second, different gates are compiled into circuits of varying depths, leading to substantial differences in their noise properties and making the characterization of $\Lambda_\mathbb{G}$ difficult to predict. For comparison, Appendix~\ref{sec:CRB} illustrates how the gate-independent noise assumption significantly simplifies the theoretical analysis of noise within the twirling group.

We now return to the discussion of the twirling group. The objective of group twirling is to regularize $\Lambda$ into a simpler form. Two widely adopted twirling groups are the Pauli group and the Clifford group, each yielding a distinct twirled channel after twirling. When the Pauli twirling group $\Pauli$ is applied, $\Lambda_\Pauli$ retains all the diagonal elements of $\Lambda$ while setting the off-diagonal elements to zero. This can be formally expressed as:
\begin{equation}
    \Lambda_\Pauli=\sum_{P\in\Pauli}\Lambda_{PP}\Pi_P,
\end{equation}
where $\Pi_P=\lket{\sigma_P}\lbra{\sigma_P}$. The twirled channel is thus fully characterized by $d^2$ parameters, known as the Pauli eigenvalues.

Twirling with the full Clifford group $\mathbb{C}_n$ results in a global depolarizing channel, which depends on only a single parameter. However, this approach generally violates the gate-independent noise assumption and incurs substantial overhead due to gate compilation~\cite{Proctor2019Direct}. To mitigate these issues, the local Clifford group $\mathbb{C}_1^{\otimes n}$ may be employed, under which the twirled channel remains diagonal and is described by only $d$ parameters. Although the Clifford group induces a more pronounced twirling effect, the Pauli group is often preferable due to its closer alignment with the practical constraints of experimental implementations.

\section{Pauli transfer character benchmarking}\label{sec:ptcb}
This section presents the PTCB protocol and demonstrates its application for benchmarking the target gate $U$ satisfying $U^2=I$. We assume that all the Pauli gates can be implemented precisely. The impact of the Pauli noise can be eliminated via the interleaved technique~\cite{magesan2012efficient}, with details provided in Appendix~\ref{sec:combination}. The target channel is modeled as:
\begin{equation}\label{eq:target noise}
    \NU=\mathcal{U}\Lambda,
\end{equation}
where $\NU$ represents the noisy implementation of the target gate, $\mathcal{U}$ is the ideal target gate, and $\Lambda$ is the noise channel.

We first establish that for any non-identity Pauli operators $P, Q$, there exists a Clifford gate $C$ such that $C P C^\dag = \pm Q$. This property follows from the fact that $H$ and $S$ gates generate the local Clifford group and enable arbitrary transpositions among ${X, Y, Z}$ (see Table~\ref{fig:conjugate}). Furthermore, CNOT gates can adjust the number of identity factors via transformations like $\textrm{CNOT}(X \otimes I)\textrm{CNOT}^\dag = X \otimes X$.
\begin{table}[!htbp]
    \centering
    \begin{tabular}{ccc}
        \hline
        $P$ & $HPH^\dag$ & $SPS^\dag$\\
        \hline
        $I$ & $I$ & $I$ \\
        $X$ & $Z$ & $Y$ \\
        $Y$ & $-Y$ & $-X$ \\
        $Z$ & $X$ & $Z$ \\
        \hline
    \end{tabular}
    \caption{Transformation of Pauli operators under conjugation by $H$ and $S$ gates.}
    \label{fig:conjugate}
\end{table}

Given a Clifford gate $C$ satisfying $C P C^\dagger = \pm Q$ (equivalently, $\mathcal{C} \lket{\sigma_P} = \pm \lket{\sigma_Q}$), the term $\NU_{PQ}$ can be moved to the diagonal by composing $\NU$ with $\mathcal{C}$. Specifically, $(\mathcal{C} \NU)_{QQ} = \NU_{PQ}$. The following circuit incorporates this technique while applying group twirling to the channel:
\begin{equation}\label{eq:full circuit}
    \mathcal{S}(P_1,P_2,\cdots,P_{2m})=\prod_{i=1}^m(\mathcal{P}_{2i}^\dag\tilde{\mathcal{U}}\mathcal{C}^\dag\mathcal{P}_{2i}\mathcal{P}_{2i-1}^\dag\mathcal{C}\tilde{\mathcal{U}}\mathcal{P}_{2i-1}).
\end{equation}
Note that the $P_1$ at the beginning, the $P_{2m}^\dag$ at the end of the circuit, and the operation between any two sequential $\NU$'s are Pauli operations. For random choices of $P_i\in\Pauli$, the expectation of the circuit is:
\begin{equation}\label{eq:ptcb expectation}
    \mathbb{E}_{P_i\in\Pauli}\mathcal{S}=((\NU\mathcal{C}^\dag)_\Pauli(\mathcal{C}\NU)_\Pauli)^m,
\end{equation}
which constitutes a diagonal matrix expressible as:
\begin{equation}\label{eq:avg}
    \mathbb{E}_{P_i\in\Pauli}\mathcal{S}=\sum_{P\in\Pauli}s_P\Pi_P.
\end{equation}
Leveraging the property $C \Pauli C^\dagger = \Pauli$, which implies that $\mathcal{C}$ is a permutation matrix with elements $\pm 1$, one can derive:
\begin{equation}
    s_Q=(\NU_{PQ}\NU_{QP})^m.
\end{equation}

To isolate the term $(\NU_{PQ} \NU_{QP})^m \Pi_Q$ from the summation in Eq.~\eqref{eq:avg} for subsequent fitting procedures, a projector can be introduced. For any $Q \in \Pauli$, there exist a set of coefficients $\lambda_P \in \{\pm 1\}$ such that~\cite{helsen2019new}:
\begin{equation}\label{eq:coefficient}
    \frac{1}{d^2}\sum_{P\in\Pauli}\lambda_P\mathcal{P}=\Pi_Q.
\end{equation}
For instance, in the single-qubit case, the projectors correspond to $\frac{1}{4}(\mathcal{I}+\mathcal{X}+\mathcal{Y}+\mathcal{Z}),\frac{1}{4}(\mathcal{I}+\mathcal{X}-\mathcal{Y}-\mathcal{Z}),\frac{1}{4}(\mathcal{I}-\mathcal{X}+\mathcal{Y}-\mathcal{Z})$, and $\frac{1}{4}(\mathcal{I}-\mathcal{X}-\mathcal{Y}+\mathcal{Z})$. Adding this weighted random Pauli gate $\lambda_P P$ at the beginning of the circuit yields:
\begin{equation}
    \mathbb{E}_{P_i\in\Pauli}\mathcal{S}\cdot \mathbb{E}_{P_0\in\Pauli}\lambda_{P_0}\mathcal{P}_0=(\NU_{PQ}\NU_{QP})^m\Pi_Q.
\end{equation}

The complete PTCB protocol is detailed in Algorithm~\ref{alg:ptcb}. The key point of our protocol is the introduction of the Clifford pair $C$ and $C^\dag$, which extracts the specific terms $\NU_{PQ}$ and $\NU_{QP}$ while ensuring that the quantum operation between the two $\NU$'s remain within the Pauli group.
\begin{algorithm}
    \caption{(PTCB protocol) Estimate $\NU_{PQ}\NU_{QP}$ for any non-identity $P,Q\in\Pauli$}
    1. Find a Clifford gate $C$ such that $CPC^\dag=\pm Q$, or equivalently, $\mathcal{C}\lket{\sigma_P}=\pm\lket{\sigma_Q}$.

    2. Let $Q=Q_1\otimes Q_2\otimes\cdots\otimes Q_n$. Prepare state $\ket{\psi_i}$, which is the $+1$ eigenstate of $Q_i$. The actual initial state $\rho$ may slightly deviate from $\ketbra{\psi_1\psi_2\cdots\psi_n}$ due to the preparation error.

    3. Uniformly sample $P_0,P_1,\cdots,P_{2m}\in\mathbb{P}$. Implement the following circuit:
    \begin{equation}
        \prod_{i=1}^m(\mathcal{P}_{2i}\tilde{\mathcal{U}}\mathcal{C}^\dag\mathcal{P}_{2i}\mathcal{P}_{2i-1}\mathcal{C}\tilde{\mathcal{U}}\mathcal{P}_{2i-1})\mathcal{P}_0.
    \end{equation}
    A more intuitive description is given in Fig.~\ref{fig:ptcb}. Measure the final state with POVM $\frac{I+Q}{2}$. This can be achieved by measuring the $i$-th qubit in the $Q_i$ basis (no measurement is performed if $Q_i = I$). The actual POVM $M$ may slightly deviate from $\frac{I+Q}{2}$ due to the measurement error. Repeat the circuit to estimate the survival probability:
    \begin{equation}
        g(m,\{P_i\})=\lbra{M}\prod_{i=1}^m(\mathcal{P}_{2i}\tilde{\mathcal{U}}\mathcal{C}^\dag\mathcal{P}_{2i}\mathcal{P}_{2i-1}\mathcal{C}\tilde{\mathcal{U}}\mathcal{P}_{2i-1})\mathcal{P}_0\lket{\rho}.
        \label{eq:survival probability2}
    \end{equation}

    4. Repeat step 3 for random sequences $\{P_i\}$. Multiply the survival probability $g(m, \{P_i\})$ by the coefficient $\lambda_{P_0}$ defined in Eq.~\eqref{eq:coefficient}, then average over all sequences to obtain:
    \begin{equation}
        \begin{aligned}
            g(m)&=\mathbb{E}_{P_i\in\mathbb{P}} \lambda_{P_0}g(m,\{P_i\})\\
            &=\lbra{M}((\NU\mathcal{C}^\dag)_\Pauli(\mathcal{C}\NU)_\Pauli)^m\Pi_Q\lket{\rho}\\
            &=\lbra{M}\Pi_Q\lket{\rho}(\NU_{PQ}\NU_{QP})^m.
        \end{aligned}
    \end{equation}
    The specific choice of initial state and POVM maximizes the overlap $\lbra{M} \Pi_Q \lket{\rho}$, thereby enhancing the signal strength and estimation accuracy.

    5. Repeat step 4 for different $m$ and implement an exponential fitting to estimate $\NU_{PQ}\NU_{QP}$. In practice, the ratio $g(1)/g(0)$ already provides a SPAM-error-free estimate.
    \label{alg:ptcb}
\end{algorithm}
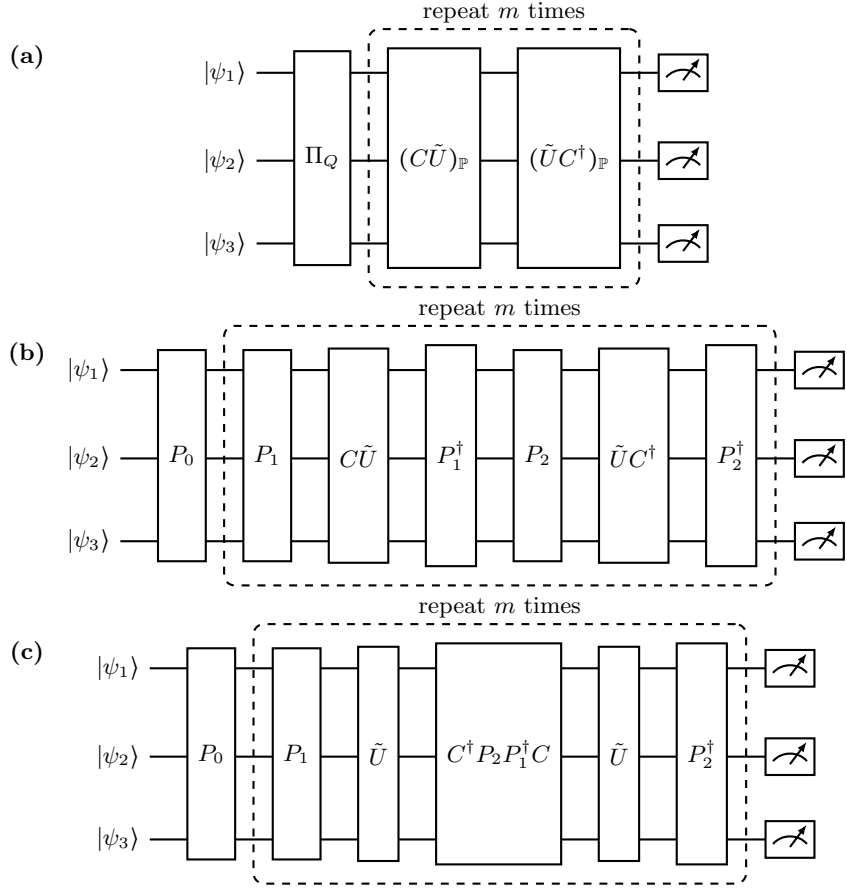
\begin{figure}[!htbp]
    \noindent
    \begin{minipage}{\textwidth}
        \begin{minipage}[t]{2em}
            \vspace{-1.5cm}
            \textbf{(a)}
        \end{minipage}%
        \begin{minipage}[t]{0.6\textwidth}
            \centering
            \begin{quantikz}
                \lstick{$\ket{\psi_1}$}&\gate[3]{\Pi_Q}&\gate[3]{(C\tilde{U})_\mathbb{P}}\gategroup[3,steps=2,style={dashed,rounded corners}]{repeat $m$ times}&\gate[3]{(\tilde{U}C^\dag)_\mathbb{P}}&\meter{}\\
            \lstick{$\ket{\psi_2}$}&&&&\meter{}\\
            \lstick{$\ket{\psi_3}$}&&&&\meter{}
            \end{quantikz}
        \end{minipage}
    \end{minipage}\\
    \begin{minipage}{\textwidth}
        \begin{minipage}[t]{2em}
            \vspace{-1.5cm}
            \textbf{(b)}
        \end{minipage}%
        \begin{minipage}[t]{0.6\textwidth}
            \centering
            \begin{quantikz}
                \lstick{$\ket{\psi_1}$}&\gate[3]{P_0}&\gate[3]{P_1}\gategroup[3,steps=6,style={dashed,rounded corners}]{repeat $m$ times}&\gate[3]{C\tilde{U}}&\gate[3]{P_1^\dag}&\gate[3]{P_2}&\gate[3]{\tilde{U}C^\dag}&\gate[3]{P_2^\dag}&\meter{}\\
            \lstick{$\ket{\psi_2}$}&&&&&&&&\meter{}\\
            \lstick{$\ket{\psi_3}$}&&&&&&&&\meter{}
            \end{quantikz}
        \end{minipage}
    \end{minipage}\\
    \begin{minipage}{\textwidth}
        \begin{minipage}[t]{2em}
            \vspace{-1.5cm}
            \textbf{(c)}
        \end{minipage}%
        \begin{minipage}[t]{0.6\textwidth}
            \centering
            \begin{quantikz}
                \lstick{$\ket{\psi_1}$}&\gate[3]{P_0}&\gate[3]{P_1}\gategroup[3,steps=5
            ,style={dashed,rounded corners}]{repeat $m$ times}&\gate[3]{\tilde{U}}&\gate[3]{C^\dag P_2P_1^\dag C}&\gate[3]{\tilde{U}}&\gate[3]{P_2^\dag}&\meter{}\\
            \lstick{$\ket{\psi_2}$}&&&&&&&\meter{}\\
            \lstick{$\ket{\psi_3}$}&&&&&&&\meter{}
            \end{quantikz}
        \end{minipage}
    \end{minipage}
    \caption{Circuit diagrams for Algorithm~\ref{alg:ptcb} (3-qubit example). The three circuits are equivalent to each other. \textbf{(a)} The expected effect of the circuit. \textbf{(b)} Group twirling with random Pauli gates. $\Pi_Q$ is also implemented by Eq.~\eqref{eq:coefficient}. \textbf{(c)} Rearrange the gates such that the operation between the two $\tilde{U}$'s remains a Pauli gate. Adjacent Pauli gates (e.g., $P_1 P_0$ and $P_3 P_2^\dagger$) should be compiled into a single Pauli operation.}
    \label{fig:ptcb}
\end{figure}

Finally, we relate the measured quantities to the fidelity. Note that:
\begin{equation}\label{eq:lambda fidelity}
    \begin{aligned}
        F(\Lambda)&=F(\mathcal{U}^\dag\NU)\\
        &=\frac{1}{d^2}\sum_{P,Q\in\Pauli}\mathcal{U}_{PQ}\NU_{PQ}\\
        &=\frac{1}{d^2}\sum_{P,Q\in\Pauli}\frac{1}{2}\mathcal{U}_{PQ}(\NU_{PQ}+\NU_{QP})\\
        &\geqslant \frac{1}{d^2}\sum_{P,Q\in\Pauli}|\mathcal{U}_{PQ}|\sqrt{\NU_{PQ}\NU_{QP}}\overset{\text{def}}{=}\hat{F}(\Lambda),
    \end{aligned}
\end{equation}
where we use the fact that $\mathcal{U}_{PQ}=\mathcal{U}_{QP}$ when $U=U^\dag$. $\hat{F}(\Lambda)$ only provides a lower bound of $F(\Lambda)$. However, for noise channel $\Lambda$ sufficiently close to the identity, the elements $\NU_{PQ}$ and $\NU_{QP}$ are nearly equal and share the sign (positive or negative) of $\mathcal{U}_{PQ}$ when $\mathcal{U}_{PQ}\neq 0$, ensuring the effectiveness of the approximation. Numerical simulations in Sec.~\ref{subsec:estimation accuracy} further validate this approach.

Since $\mathcal{U}$ and $\NU$ are trace-preserving maps, we have $\mathcal{U}_{IP}=\mathcal{U}_{PI}=0$ for $P\neq I$, and $\mathcal{U}_{II}=\NU_{II}=1$. Although Algorithm~\ref{alg:ptcb} is restricted to $P,Q\neq I$, $\hat{F}(\Lambda)$ can still be evaluated for all $(P, Q)$ pairs with $\mathcal{U}_{PQ} \neq 0$. If $\mathcal{U}$ contains an impractically large number of non-zero elements, randomly sampling a subset of $(P, Q)$ pairs provides a sufficient estimate for $F(\Lambda)$.

Moreover, the requirement $U=U^\dag$ can be relaxed under the assumption that the noisy implementation of the inverse gate satisfies $\widetilde{\mathcal{U}^\dag}=(\NU)^\dag$, i.e., the PTM of noisy $U^\dag$ is the conjugate transpose of the PTM of noisy $U$. To leverage this, we modify the circuit in Eq.~\eqref{eq:full circuit} by replacing a $U$ operation with $U^\dag$ operation:
\begin{equation}
    \begin{gathered}
        \mathcal{S}(P_1,P_2,\cdots,P_{2m})=\prod_{i=1}^m(\mathcal{P}_{2i}^\dag\widetilde{\mathcal{U}^\dag}\mathcal{C}^\dag\mathcal{P}_{2i}\mathcal{P}_{2i-1}^\dag\mathcal{C}\tilde{\mathcal{U}}\mathcal{P}_{2i-1}),\\
        \mathbb{E}_{P_i\in\Pauli}\mathcal{S}=((\widetilde{\mathcal{U}^\dag}\mathcal{C}^\dag)_\Pauli(\mathcal{C}\NU)_\Pauli)^m.
    \end{gathered}
\end{equation}
With all other operations remaining unchanged, Algorithm~\ref{alg:ptcb} now yields $\NU_{PQ}(\widetilde{\mathcal{U}^\dag})_{QP}=\NU_{PQ}^2$. This enables direct estimation of $F(\Lambda)$, without relying on the inequality in Eq.~\eqref{eq:lambda fidelity}, thereby extending the protocol's applicability beyond self-adjoint gates.

\section{Numerical simulation}\label{sec:simulation}
This section presents the numerical simulation results of benchmarking Toffoli gates using the PTCB protocol, along with a detailed description of the techniques and parameters employed. For convenience, we define the infidelity as $1 - \text{fidelity}$.

The benchmarking procedure for Toffoli gates involves three levels of randomness. At the highest level, multiple pairs of $(P,  Q)$ are randomly selected to estimate $\hat{F}(\Lambda)$ (as defined in Eq.~\eqref{eq:lambda fidelity}), termed \textit{outer sampling}. At the intermediate level, multiple sets of ${P_i}$ are randomly chosen to estimate $g(m)$ (Step 4 of Algorithm~\ref{alg:ptcb}), termed \textit{inner sampling}. At the lowest level, each circuit configuration is repeated multiple times to estimate the survival probability (Step 3 of Algorithm~\ref{alg:ptcb}), termed \textit{circuit repetition}. This section first introduces the noise model and subsequently analyzes the sample sizes required at each stage to ensure estimation accuracy. For simplicity, the noise of Pauli twirling gates is neglected.
\subsection{Noise model}
The noise channel is constructed from three components: local dephasing, local amplitude damping, and unitary noise. We choose the same noise model as in~\cite{liu2024group}, because this model simultaneously incorporates both Pauli noise and non-Pauli noise. Specifically, the strength parameters for each noise type are randomly sampled from specified intervals, and the final noise channel is formed by composing the corresponding PTMs. The single-qubit dephasing and amplitude damping channels are defined as:
\begin{equation}
    \begin{gathered}
        \Lambda_{\textrm{dephase}}(\rho)=K_0\rho K_0^\dag+K_1\rho K_1^\dag,\\
        \Lambda_{\textrm{damp}}(\rho)=K_2\rho K_2^\dag+K_3\rho K_3^\dag,
    \end{gathered}
\end{equation}
where the Kraus operators are given by:
\begin{equation}
    \begin{gathered}
        K_0=\begin{pmatrix}
            \sqrt{1-p}&0\\
            0&\sqrt{1-p}
        \end{pmatrix},\
        K_1=\begin{pmatrix}
            \sqrt{p}&0\\
            0&-\sqrt{p}
        \end{pmatrix},\\
        K_2=\begin{pmatrix}
            1&0\\
            0&\sqrt{1-q}
        \end{pmatrix},\
        K_3=\begin{pmatrix}
            0&\sqrt{q}\\
            0&0
        \end{pmatrix}.
    \end{gathered}
\end{equation}
Here, $p$ and $q$ parameterize the noise strength. We assume identical noise parameters for each qubit, resulting in global noise channels $\Lambda_{\textrm{dephase}}^{\otimes3}$ and $\Lambda_{\textrm{damp}}^{\otimes3}$.

We model the unitary noise as controlled-$\exp(i\delta X)$ gates:
\begin{equation}
    \begin{gathered}
        \Lambda_{\textrm{unitary}}(\rho)=V\rho V^\dag,\\
        V=\ketbra{0}\otimes I+\ketbra{1}\otimes e^{i\delta X}.
    \end{gathered}
\end{equation}
We adopt this modeling because the CNOT gate is a crucial component of the Toffoli gate implementation and is also the primary source of errors. The control qubit, target qubit, and noise strength $\delta$ are all randomly sampled. The composite noise channel is $\Lambda = \Lambda_{\textrm{dephase}} \circ \Lambda_{\textrm{unitary}} \circ \Lambda_{\textrm{damp}}$. By varying the noise parameters, we generate an ensemble of noise channels whose infidelities are uniformly distributed within the interval $[0.01, 0.04]$.

State preparation errors are modeled by assuming the initial state $\ketbra{0}$ is incorrectly prepared as $\ketbra{1}$ with a fixed probability $p$, yielding the actual initial state $(1-p)\ketbra{0} + p\ketbra{1}$. Measurement errors are modeled by assuming each measurement outcome is flipped with probability $p$. That is, for ideal POVM elements $P_1$ and $P_2$, the realized POVMs are $(1-p)P_1 + p P_2$ and $(1-p)P_2 + p P_1$, respectively.

\subsection{Accuracy of the estimation formula}\label{subsec:estimation accuracy}
\begin{figure}[!htbp]
    \includegraphics[scale=0.4]{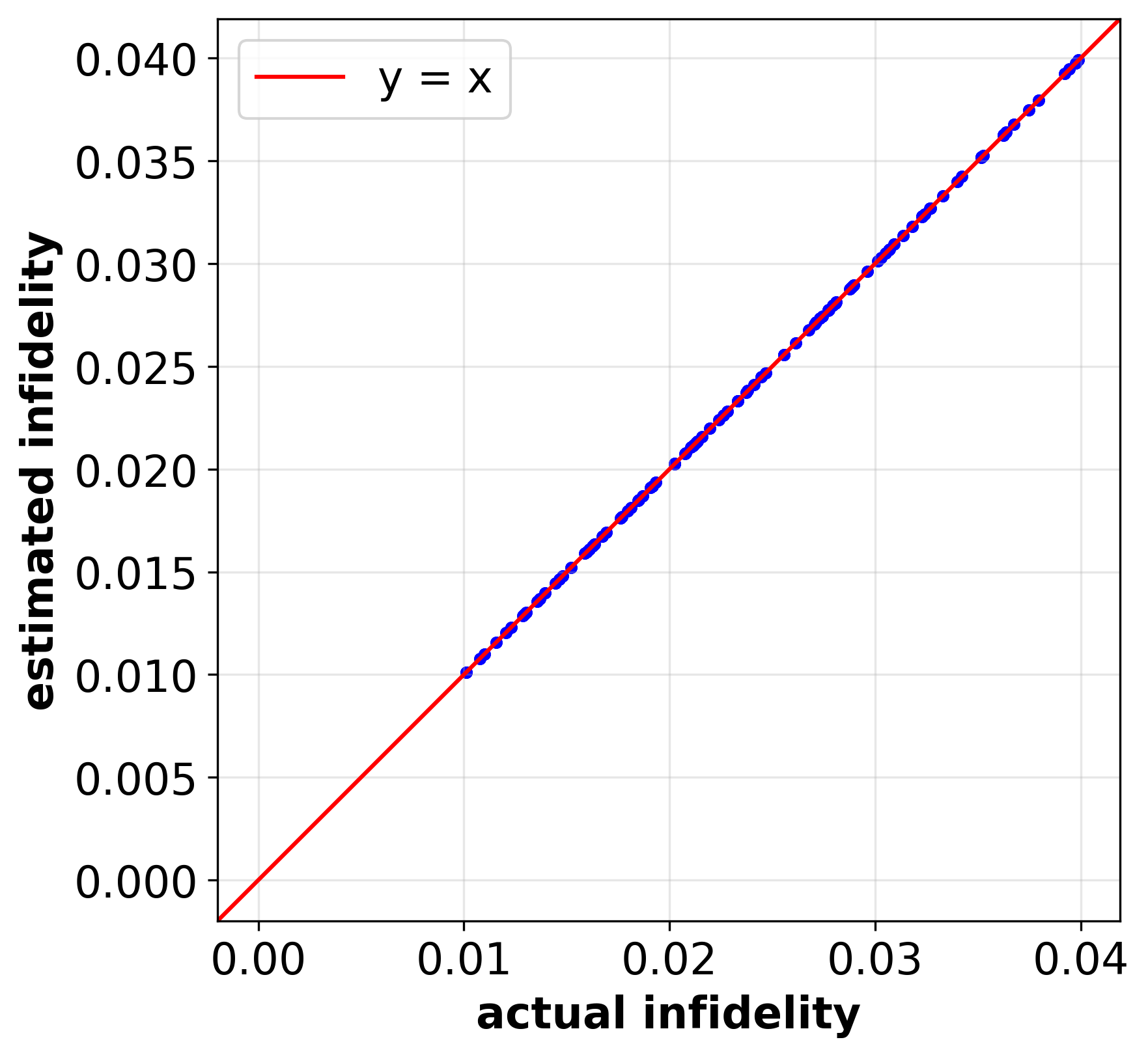}
    \hspace{2em}
    \includegraphics[scale=0.4]{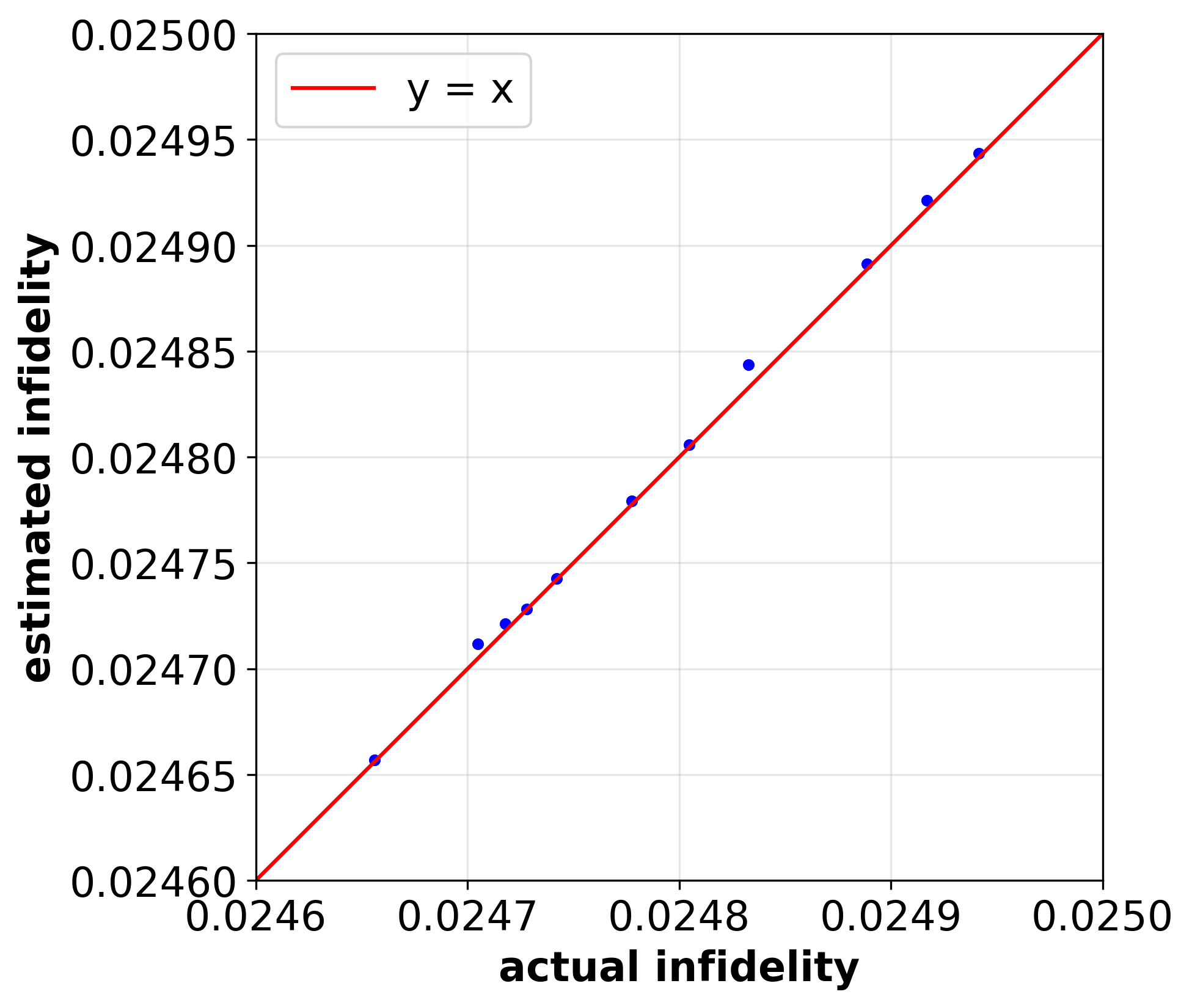}
    \caption{Discrepancy between the actual infidelity $F(\Lambda)$ and the estimated infidelity $\hat{F}(\Lambda)$. The right figure is an enlarged version for the region between 0.0246 and 0.0250 within the left figure. For all simulated noise channels, the discrepancy remains below $1 \times 10^{-4}$.}
    \label{fig:estimation}
\end{figure}

Recall that $\hat{F}(\Lambda)$ provides a lower bound for $F(\Lambda)$. Figure~\ref{fig:estimation} shows the discrepancy between the actual infidelity $F(\Lambda)$ and the estimated infidelity $\hat{F}(\Lambda)$, demonstrating that the bias introduced by this approximation is negligible. For all simulated noise channels, the discrepancy does not exceed $1 \times 10^{-4}$.
\subsection{Outer sampling}
In this subsection, we disregard the randomness from inner sampling and circuit repetition, assuming exact knowledge of $\NU_{PQ}\NU_{QP}$. A technique of importance sampling can facilitate faster convergence of the estimation toward $F(\Lambda)$, as detailed below. Note that:
\begin{equation}
    \begin{aligned}
        F(\Lambda)&\geqslant\frac{1}{d^2}\sum_{P,Q\in\Pauli}|\mathcal{U}_{PQ}|\sqrt{\NU_{PQ}\NU_{QP}}\\
        &=\frac{1}{d^2}\sum_{P,Q\in\Pauli}\mathcal{U}_{PQ}^2\frac{\sqrt{\NU_{PQ}\NU_{QP}}}{|\mathcal{U}_{PQ}|}\\
        &=\mathbb{E}_{P,Q}\frac{\sqrt{\NU_{PQ}\NU_{QP}}}{|\mathcal{U}_{PQ}|},
    \end{aligned}
\end{equation}
where the expectation is taken over $(P,Q)$ sampled with probability $\frac{1}{d^2}\mathcal{U}_{PQ}^2$. The PTM $\mathcal{U}$ for the Toffoli gate contains $232$ non-zero entries: $8$ entries equal to $1$, and $224$ entries equal to $\pm \frac{1}{2}$. We partition the total probability into $256$ equal segments. Each $(P, Q)$ pair with $\mathcal{U}_{PQ} = 1$ occupies $4$ segments, while each pair with $\mathcal{U}_{PQ} = \pm\frac{1}{2}$ occupies $1$ segment. Let $M$ denote the number of outer samples. The sampling procedure reduces to selecting $M$ distinct numbers from $1$ to $256$. The estimate is then
\begin{equation}\label{eq:outer}
    \frac{1}{M}\sum_{i=1}^{M}\frac{\sqrt{\NU_{P_iQ_i}\NU_{Q_iP_i}}}{|\mathcal{U}_{P_iQ_i}|}.
\end{equation}
Since each term in the summation is expected to be close to $1$, this method promotes faster convergence.

\begin{figure}[!htbp]
    \includegraphics[scale=0.4]{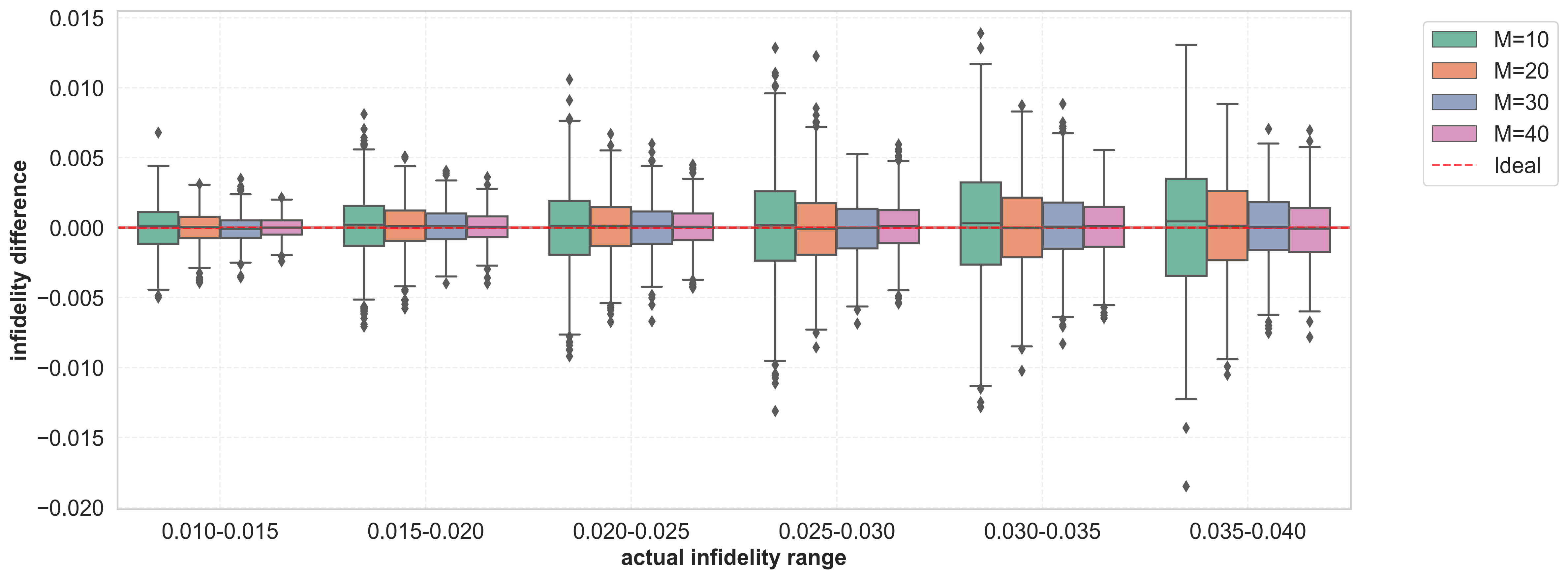}
    \caption{Discrepancy between the actual infidelity and the infidelity estimated via Eq.~\eqref{eq:outer} for different values of $M$. Results are grouped by actual infidelity to demonstrate protocol performance across different noise levels. As $M$ increases, the estimated infidelity converges toward the actual value. Larger actual infidelity is accompanied by increased estimation variance.}
    \label{fig:outer}
\end{figure}

Figure~\ref{fig:outer} displays the discrepancy between the actual and estimated infidelity for varying $M$. Balancing accuracy and computational cost, we select $M = 30$ for the analysis in Sec.~\ref{subsec:inner}.
\subsection{Inner sample and circuit repetition}\label{subsec:inner}
We utilize the ratio $g(1)/g(0)$ to estimate $\NU_{PQ}\NU_{QP}$ (Step 5 of Algorithm~\ref{alg:ptcb}). The estimation of $g(0)$ involves sampling only $P_0$, which has $64$ possible choices. Exhaustive enumeration of all $P_0$ is computationally feasible compared to the random sampling required for $g(1)$. Furthermore, since $g(0)$ appears in the denominator, any estimation error can significantly amplify the overall error. Thus, obtaining an accurate value for $g(0)$ is crucial. The estimation of $g(1)$ involves sampling $P_0,P_1,P_2$. We randomly sample $M'$ distinct sets of $(P_0,P_1,P_2)$ to estimate $g(1)$.

\begin{figure}[!htbp]
    \noindent
    \begin{minipage}{\textwidth}
        \begin{minipage}[t]{2em}
            \vspace{-5.6cm}
            \textbf{(a)}
        \end{minipage}%
        \begin{minipage}[t]{0.9\textwidth}
            \includegraphics[scale=0.4]{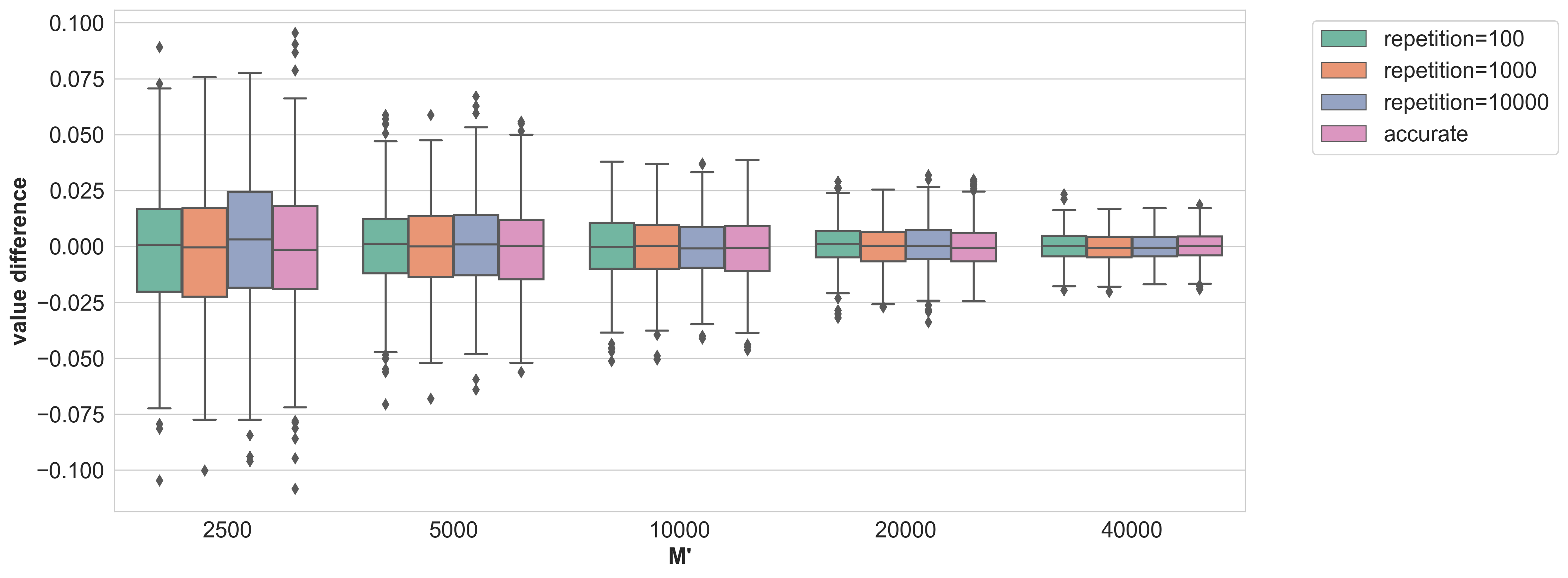}
        \end{minipage}
    \end{minipage}\\
    \begin{minipage}{\textwidth}
        \begin{minipage}[t]{2em}
            \vspace{-5.6cm}
            \textbf{(b)}
        \end{minipage}%
        \begin{minipage}[t]{0.9\textwidth}
            \includegraphics[scale=0.4]{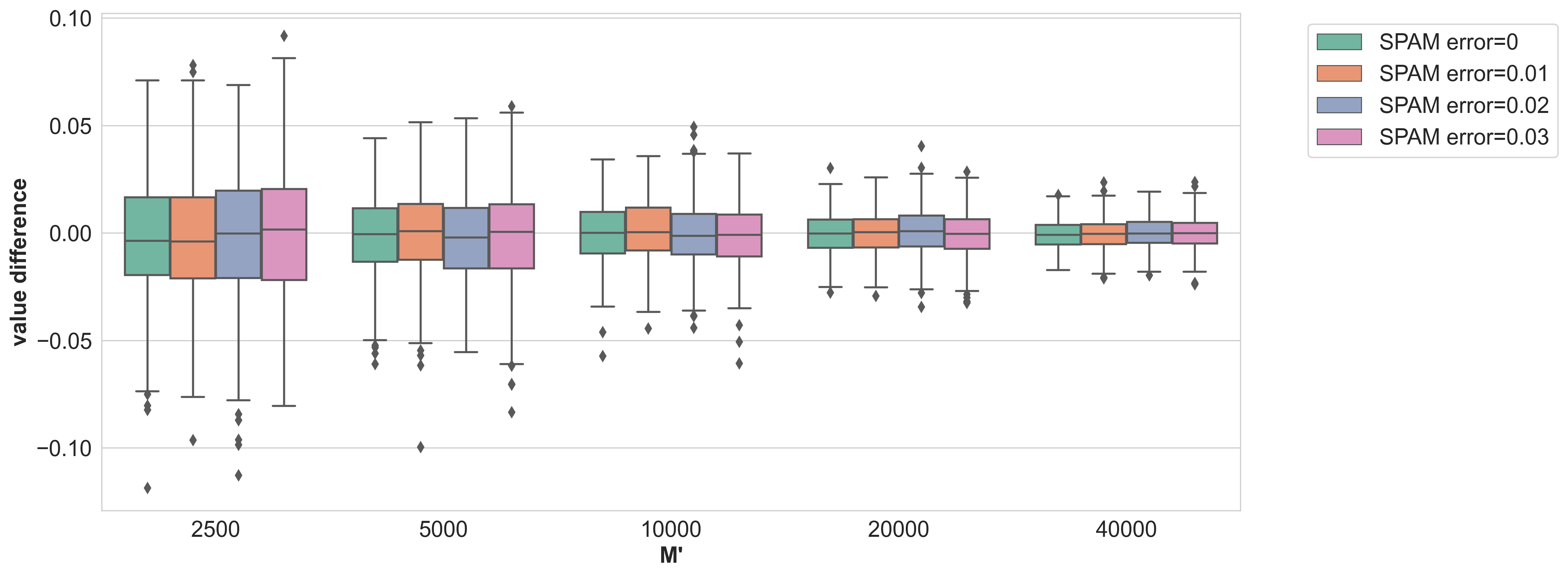}
        \end{minipage}
    \end{minipage}
    \caption{Discrepancy between $\sqrt{\NU_{PQ}\NU_{QP}}$ and the square root of the value estimated by PTCB protocol. \textbf{(a)} Impact of repetition count at a fixed SPAM error rate. The label ``accurate” denotes direct acquisition of the survival probability without circuit repetition. As $M'$ increases, the estimate converges to the actual value, while the repetition count has negligible impact. \textbf{(b)} Impact of SPAM error rate at a fixed repetition count. As $M'$ increases, the estimate converges to the actual value, while the SPAM error rate has little impact, demonstrating the SPAM-error-free nature of the protocol.}
    \label{fig:rep}
\end{figure}

\begin{figure}[!htbp]
    \includegraphics[scale=0.4]{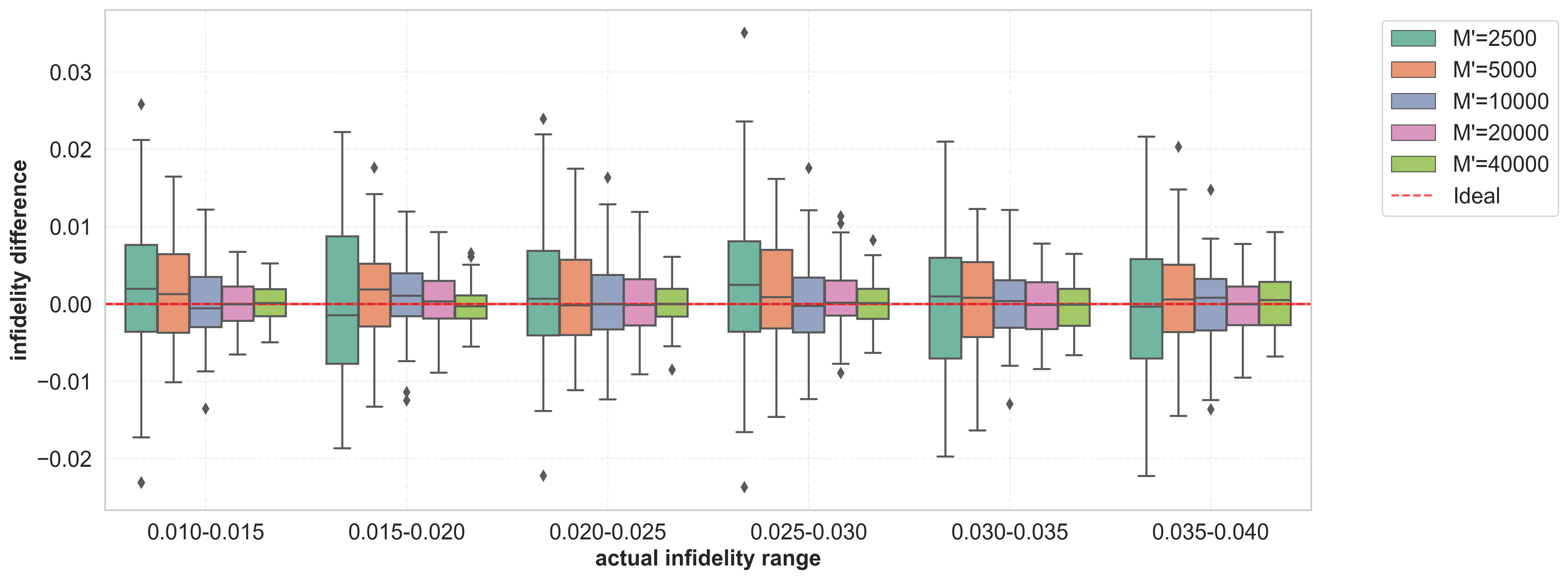}
    \caption{Discrepancy between the actual infidelity and the estimated infidelity with $M=30$. As $M'$ increases, the estimated infidelity converges to the actual value. The actual infidelity level has little impact on the estimation fluctuation.}
    \label{fig:inner}
\end{figure}

We first select a single $(P,Q)$ pair with $U_{PQ}=\frac{1}{2}$ (specifically, $P=IIY,\ Q=IZY$) and execute the PTCB protocol for different values of $M'$, repetition counts, and SPAM error rates. This choice is motivated by two factors: firstly, ``$\pm\frac{1}{2}$" terms constitute the majority of the probability; secondly, ``$1$" terms lie on the diagonal, while PTCB and character RB exhibit no essential difference when benchmarking diagonal elements. The results, shown in Fig.~\ref{fig:rep}, indicate that the number of repetitions has minimal impact on the estimation, and the protocol remains robust to SPAM errors. Therefore, for simplicity, we assume that the survival probability is obtained directly without circuit repetition when examining the impact of $M'$ on the estimation. The results are presented in Fig.~\ref{fig:inner}.

\section{Discussion}\label{sec:discussion}
This work proposes the PTCB protocol, a SPAM-error-free method that utilizes only the Pauli twirling group and Pauli-basis SPAM to estimate the product of any two symmetric elements of the PTM for a quantum channel. Furthermore, we develop a benchmarking protocol for unitary gates $U$ satisfying $U^2=I$ (a constraint that can be relaxed under appropriate conditions) by incorporating PTCB as a core module. This approach establishes a theoretical pathway for benchmarking non-Clifford gates using exclusively local operations, a capability of significant importance for the development of fault-tolerant quantum computing.

Several promising directions for future research remain open. First, while the PTCB protocol is conceptually general, its scaling behavior with respect to the number of qubits warrants further investigation and potential optimization. For instance, the PTM of the multi-qubit controlled $X$ gate ($C^nX$) contains terms that decay exponentially. Accurate estimation of these terms under the current framework would necessitate a large sampling to achieve convergence within a desired error tolerance. Second, our numerical simulations indicate that the current PTCB protocol exhibits non-negligible variance in estimating the fidelity of Toffoli gates, which suggests room for optimization toward practical near-term applications. Future work could focus on developing more efficient variants or complementary techniques.

To address the latter challenge, we identify two potential avenues for enhancement. One direction involves exploring alternative twirling groups, such as the local Clifford group, which may induce a stronger twirling effect and further simplify the channel's structure. Integrating such groups within the PTCB framework could reduce the number of parameters and thereby decrease the sampling overhead. Another direction is to consider alternative fidelity metrics better suited to the protocol’s features. For example, with prior knowledge about the noise channel, one could identify which values of $\NU_{PQ}\NU_{QP}$ are most sensitive to errors and focus resources on optimizing those specific terms, potentially leading to more efficient benchmarking strategies.

\begin{acknowledgements}
This work was supported by the National Natural Science Foundation of China Grant No.~12174216 and the Innovation Program for Quantum Science and Technology Grant No.~2021ZD0300804 and No.~2021ZD0300702.
\end{acknowledgements}

\appendix

\section{Character randomized benchmarking}\label{sec:CRB}
This section introduces the character RB protocol~\cite{helsen2019new}, which serves as the foundation for the PTCB protocol. Character RB provides a method for estimating the fidelity of a group of gates. Here, we focus on its application to the Pauli group, which is a prerequisite step for separating the Pauli gate noise from the combined noise discussed in Appendix~\ref{sec:combination}.

Since any multi-qubit Pauli gate is composed of a single layer of single-qubit gates, it is reasonable to assume that all the Pauli gates share a common noise channel. This is called the gate-independent noise assumption, and can be expressed as:
\begin{equation}\label{eq:pauli noise}
    \tilde{\mathcal{P}}=\mathcal{E}\mathcal{P},
\end{equation}
where $\tilde{\mathcal{P}}$ represents the actually implemented channel, $\mathcal{P}$ is the ideal Pauli channel and $\mathcal{E}$ denotes the noise channel. Note that the order of the ideal gate and noise channel differs from Eq.~\eqref{eq:target noise}; however, this is inconsequential since $\mathcal{EP} = \mathcal{P}(\mathcal{P}^\dagger\mathcal{EP})$, allowing $\mathcal{P}^\dagger\mathcal{EP}$ to be interpreted as a modified noise channel. The following circuit implements twirling of the noise channel:
\begin{equation}
    \begin{aligned}
        \mathcal{S}(P_1,P_2,\cdots,P_m)&=\widetilde{\mathcal{P}_{m}^\dag}(\prod_{i=1}^{m-1}\widetilde{\mathcal{P}_{i+1}\mathcal{P}_i^\dag})\widetilde{\mathcal{P}_1}\\
        &=\mathcal{EP}^\dag_m(\prod_{i=1}^{m-1}\mathcal{E}\mathcal{P}_{i+1}\mathcal{P}_i^\dag)\mathcal{EP}_1\\
        &=\mathcal{E}\prod_{i=1}^{m}\mathcal{P}_i^\dag\mathcal{EP}_i,
    \end{aligned}
\end{equation}
where $m$ is a parameter controlling the depth of the circuit. Note that $P_{i+1}P_i^\dag$ remains a Pauli gate and should be implemented as a single operation with only one associated noise channel. For random choices of $P_i\in\Pauli$, the expectation of the circuit is
\begin{equation}
    \mathbb{E}_{P_i\in\Pauli}\mathcal{S}=\mathcal{E}\mathcal{E}_\Pauli^m=\mathcal{E}\sum_{P\in\Pauli}\mathcal{E}_{PP}^m\Pi_P.
\end{equation}
This construction successfully generates the $\mathcal{E}{PP}^m$ terms (exponents of the Pauli eigenvalues), but they appear in a summation. To isolate individual Pauli eigenvalues, we employ the projector technique introduced in Eq.~\eqref{eq:coefficient}:
\begin{equation}
    \mathbb{E}_{P_i\in\Pauli}\mathcal{S}\cdot \mathbb{E}_{P_0\in\Pauli}\lambda_{P_0}\mathcal{P}_0= \mathcal{E}\mathcal{E}_{QQ}^m\Pi_Q.
\end{equation}
Here, the noise channel for $P_0$ is omitted since $P_1P_0$ is implemented as a single Pauli gate. The complete algorithm is detailed in Algorithm~\ref{alg:crb}.

\begin{algorithm}
    \caption{Estimate $\mathcal{E}_{QQ}$ for any $Q\in\Pauli$}
    1. Let $Q=Q_1\otimes Q_2\otimes\cdots\otimes Q_n$. Prepare state $\ket{\psi_i}$, which is the $+1$ eigenvalue of $Q_i$. The actual initial state $\rho$ may slightly deviate from $\ketbra{\psi_1\psi_2\cdots\psi_n}$ due to the preparation error.

    2. Uniformly sample $P_0,P_1,\cdots,P_{m}\in\Pauli$. Implement the following circuit,
    \begin{equation}
        \widetilde{\mathcal{P}_{m}^\dag}(\prod_{i=1}^{m-1}\widetilde{\mathcal{P}_{i+1}\mathcal{P}_i^\dag})\widetilde{\mathcal{P}_1\mathcal{P}_0}.
    \end{equation}
    A more intuitive description is given in Fig.~\ref{fig:crb}. Measure the final state with POVM $\frac{I+Q}{2}$. This can be achieved by measuring the $i$-th qubit in the $Q_i$ basis (no measurement is performed if $Q_i=I$). The actual POVM $M$ may slightly deviate from $\frac{I+Q}{2}$ due to the measurement error. Repeat the circuit to estimate the survival probability:
    \begin{equation}
        f(m,\{P_i\})=\lbra{M}\prod_{i=1}^m\widetilde{\mathcal{P}_{m}^\dag}(\prod_{i=1}^{m-1}\widetilde{\mathcal{P}_{i+1}\mathcal{P}_i^\dag})\widetilde{\mathcal{P}_1\mathcal{P}_0}\lket{\rho}
        \label{eq:survival probability}
    \end{equation}

    3. Repeat step 2 for randomly sequences $\{P_i\}$. Multiply the survival probability $f(m,\{P_i\})$ by the coefficient $\lambda_{P_0}$ defined in Eq.~\eqref{eq:coefficient}, then average over all sequences to obtain:
    \begin{equation}
        \begin{aligned}
            f(m)&=\mathbb{E}_{P_i\in\mathbb{P}} \lambda_{P_0}f(m,\{P_i\})\\
            &=\lbra{M}\mathcal{E}_{QQ}^m\mathcal{E}\Pi_Q\lket{\rho}\\
            &=\lbra{M'}\Pi_Q\lket{\rho}\mathcal{E}_{QQ}^m,
        \end{aligned}
    \end{equation}
    where $\lbra{M'}=\lbra{M}\mathcal{E}$, indicating that $\mathcal{E}$ is absorbed into measurement error. The specific choice of initial state and POVM maximize the overlap $\lbra{M'}\Pi_Q\lket{\rho}$, thereby enhancing the signal strength and estimation accuracy.

    4. Repeat step 3 for different $m$ and implement an exponential fitting to estimate $\mathcal{E}_{QQ}$. For instance, apply least squares regression to $\ln f(m)$ versus $m$.
    \label{alg:crb}
\end{algorithm}
\begin{figure}[!htbp]
    \begin{quantikz}
        \lstick{$\ket{\psi_1}$}&\gate[3]{P_1P_0}&\gate[3]{P_2P_1^\dag}&\gate[3]{P_3P_2^\dag}&\qw\ \ldots\ &\gate[3]{P_mP_{m-1}^\dag}&\gate[3]{P_m^\dag}&\meter{}\\
        \lstick{$\ket{\psi_2}$}&&&&\qw\ \ldots\ &&&\meter{}\\
        \lstick{$\ket{\psi_3}$}&&&&\qw\ \ldots\ &&&\meter{}
    \end{quantikz}
    \caption{Circuit diagram for Algorithm~\ref{alg:crb} (3-qubit example). Each Block represents a single Pauli gate implementation, so the noise channel is introduced only once per operation.}
    \label{fig:crb}
\end{figure}
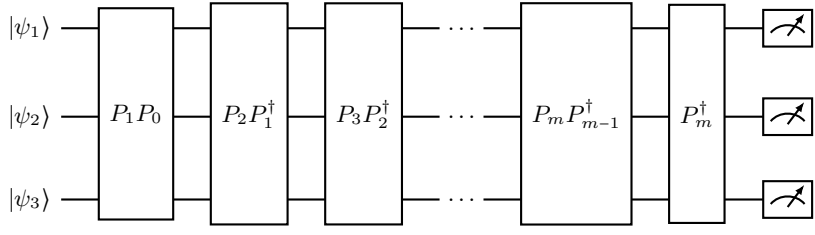

Finally, recall that the fidelity can be expressed as:
\begin{equation}
    F(\mathcal{E})=\frac{1}{d^2}\sum_{Q\in\Pauli}\mathcal{E}_{QQ}.
\end{equation}
This summation involves $d^2$ sum terms, making exhaustive estimation of all $\mathcal{E}_{QQ}$ impractical. In practice, one can either randomly sample a subset of $Q$ values to estimate $F(\mathcal{E})$, or employ the character average benchmarking \cite{zhang2022scalable}.

\section{Combination of the target noise and the Pauli gate noise}\label{sec:combination}
In Sec.~\ref{sec:ptcb}, the Pauli noise was neglected. When this noise is taken into consideration, Eq.~\eqref{eq:ptcb expectation} should be modified as follows:
\begin{equation}
    \mathbb{E}_{P_i\in\Pauli}\mathcal{S}=\mathcal{E}((\NU\mathcal{E}\mathcal{C}^\dag)_\Pauli(\mathcal{C}\NU\mathcal{E})_\Pauli)^m.
\end{equation}
In this scenario, Algorithm~\ref{alg:ptcb} yields the product $(\NU\mathcal{E})_{PQ}(\NU\mathcal{E})_{QP}$, and Eq.~\eqref{eq:lambda fidelity} consequently provides an estimate of:
\begin{equation}
    \begin{aligned}
        &\frac{1}{d^2}\sum_{P,Q\in\Pauli}|\mathcal{U}_{PQ}|\sqrt{(\NU\mathcal{E})_{PQ}(\NU\mathcal{E})_{QP}}\\
        \approx&\frac{1}{d^2}\sum_{P,Q\in\Pauli}\mathcal{U}_{PQ}(\NU\mathcal{E})_{PQ}\\
        =&F(\mathcal{U}^\dag\NU\mathcal{E})\\
        =&F(\Lambda\mathcal{E}).
    \end{aligned}
\end{equation}
A technique known as interleaved RB for estimating $F(\Lambda)$ from $F(\Lambda\mathcal{E})$ and $F(\mathcal{E})$ has been introduced \cite{magesan2012efficient,liu2024group}. We state the result directly: $F(\Lambda)$ lies within the interval
\begin{equation}
    [\frac{d^2(F(\Lambda\mathcal{E})-E)-1}{d^2F(\mathcal{E})-1}(1-\frac{1}{d^2})+\frac{1}{d^2},\frac{d^2(F(\Lambda\mathcal{E})+E)-1}{d^2F(\mathcal{E})-1}(1-\frac{1}{d^2})+\frac{1}{d^2}],
\end{equation}
where $E$ can be upper bounded by:
\begin{equation}
    E\leqslant\min\begin{cases}
        4(d+1)\sqrt{1-F(\mathcal{E})}+2\frac{d+1}{d}(1-F(\mathcal{E}))\\
        \frac{|d^2(F(\Lambda\mathcal{E})-F(\mathcal{E}))+2F(\mathcal{E})-F(\Lambda\mathcal{E})-1|+(d^2F(\mathcal{E})-1)(1-F(\mathcal{E}))}{d^2-1}\\
        \frac{|d^2(F(\Lambda\mathcal{E})-F(\mathcal{E}))+2F(\mathcal{E})-F(\Lambda\mathcal{E})-1|+(d^2F(\mathcal{E})-1)(F(\mathcal{E})-F(\Lambda\mathcal{E}))}{d^2-1}
    \end{cases}
\end{equation}

It is evident that as $F(\mathcal{E})$ approaches $1$, the term $E$ tends toward $0$, leading to a more precise estimation of $F(\Lambda)$. This observation underscores the advantage of employing the Pauli twirling group, given that most current quantum hardware platforms can implement Pauli gates with relatively high fidelity.

\bibliographystyle{apsrev4-1}

\bibliography{bibCCZ}

\end{document}